\documentclass[pre,unsortedaddress,superscriptaddress,showkeys]{revtex4}
\usepackage{graphicx} 
\DeclareGraphicsExtensions{.eps}
\setkeys{Gin}{width=\columnwidth}

\usepackage{amsmath}

\newcommand{\consortium}{\affiliation{Consortium of the Americas for
Interdisciplinary Science and Department of
Physics and Astronomy, \mbox{University of New Mexico,
Albuquerque, New Mexico 87131}}}

\begin{document}


\title{A Theoretical Framework for the Analysis
of the \\West
Nile Virus Epidemic}

\author{V. M. Kenkre}
\thanks{corresponding author}
\consortium

\author{R. R. Parmenter}
\affiliation{Department of Biology, University of
New Mexico, Albuquerque, New Mexico 87131}

\author{I. D. Peixoto}
\consortium
\affiliation{Instituto Balseiro and Centro
At\'omico Bariloche, 8400 San Carlos de Bariloche,
Argentina}

\author{L. Sadasiv}
\consortium

\begin{abstract}
We present a model for the growth of West Nile
virus in mosquito and bird populations based on
observations of the initial epidemic in the U.S.
Increase of bird mortality as a result of infection, 
which is a feature of the epidemic, is found to yield 
an effect which is observable in principle, viz.,  periodic 
variations in the extent of infection. The vast difference 
between mosquito and bird lifespans, another peculiarity 
of the system, is shown to lead to interesting consequences
regarding delay in the onset of the steady-state infection. An
outline of a framework is provided to treat mosquito diffusion and bird migration.
\end{abstract}

\date{\today} \maketitle

\section{Introduction}

Mathematical modeling of the spread of epidemics
poses intriguing challenges and promises useful
insights and possibly predictive capabilities.
Recent work by some of the present coauthors
\cite{ak,aakb,akyp} has led to the understanding
of observed features, particularly spatiotemporal
patterns, in the Hantavirus infection
\cite{yates}. It is the purpose of the present
paper to initiate a formalism for the
understanding of the West Nile virus epidemic,
which bears some similarities, but possesses some
distinguishing characteristics, relative to the
Hantavirus. The paper is laid out as follows.  In
the rest of the section we describe some essential
characteristics of the West Nile virus epidemic and
comment on how they may be folded into a model of
differential equations similar to the
Abramson-Kenkre (AK) model of the Hantavirus
\cite{ak}. In Section \ref{gen_model}, we modify
the Hantavirus model equations to incorporate
cross-infection of two taxa, a characteristic
of the West Nile virus epidemic. We comment on general
features expected on the grounds of simple
intuition based on nonlinear dynamics. In Section
\ref{realistic_model}, we augment the generalized
cross-infection model to include three realistic
features of the West Nile virus epidemic and present
scenarios for time evolution of the populations of
mosquitoes and birds, the two taxa which appear
central to the West Nile virus problem. The present
paucity of field data prevents us from attempting
to explain specific observations. However,
interesting predictions of the `what-if' type are
possible as will be seen below. Concluding remarks
form Section \ref{conclusion}.

West Nile virus is a mosquito-borne virus that
infects primarily birds, but also a wide range of other species, including horses, dogs and cats, and
occasionally humans. The first outbreak of West
Nile virus encephalitis on the North American
continent occurred in New York in 1999.
Successive outbreaks in humans have occurred
annually in the USA since then.  West Nile virus is
fatal in many species of birds, and is sometimes
fatal in humans.  It is unusual among
mosquito-borne diseases in that ``vertical
transmission'', where the virus is passed from
the mother to her eggs, may occur in the wild.
This has potentially serious consequences, because
once an area is infected it may remain so
indefinitely, because the virus may survive the
winter in infected mosquito larvae and reemerge to infect
human and animal populations in the spring
\cite{Nasci}.

Previous work has shown that the virus travels
along watershed areas through avian and mosquito
host populations \cite{MarfinCID}.  Extensive
field studies have led to attention being focused
on birds as well as mosquitoes in the dynamics of
the West Nile virus epidemic. During their migration,
infected birds arrive at a location and transmit
the virus to female mosquitoes that feed upon them. The
mosquitoes in turn transmit the virus to other
birds, not originally infected, and to other
animals including horses and humans. Collection of
field data consists, therefore, of testing
mosquitoes and birds. Mosquitoes are trapped with
CO${}_2$-releasing boxes with organic-rich water
at their base.  In addition, surveillance systems
for reporting dead birds and testing them for
infection, as well as trapping live birds and
testing them for seroconversion (a symptom of
recent West Nile virus infection) are in place in
centers of infection
\cite{MarfinEID,Theophilides}.

The model that we develop for studies of the West Nile virus
epidemic is similar to the AK analysis of the
Hantavirus \cite{ak} but incorporates the above as
important additional features. From the point of
view of modeling, three additions to the AK model
are crucial. The first is that there are
\emph{two} taxa in the West Nile virus system (mosquitoes
and birds, as compared to the single rodent species
in the Hantavirus), that cross-infect each other. The
second is that these two taxa have vastly
different (natural) lifespans: the characteristic
times are on the order of a few weeks for mosquitoes
but on the order of a year or two for birds. The
third is that while mice are never born infected with Hantavirus,
mosquitoes may be hatched West Nile virus; and while mice do not die from
hanta infection, birds often do die from West Nile virus
infection. The first addition is treated in
Section~\ref{gen_model} where we find obtain the generalized condition for 
steady state  infection to exist for a system with
two populations.  The second addition is shown in
Section~\ref{realistic_model} to lead to interesting consequences in the onset of the
steady-state infection. The treatment of the third
addition, also in Section~\ref{realistic_model},
shows that the increase in bird mortality due to
infection can lead to oscillations in infected
populations.

\section{Generalization of the Hantavirus Model to
Include Cross-Infection} \label{gen_model}

\subsection{Recasting the Hantavirus
Equations}\label{recasting}

The AK model equations, at the mean-field level
(at which diffusion is not shown explicitly), are
\begin{subequations} \label{eq:ak} \begin{align}
\frac{dm_s}{dt} &= b\,m - c\,m_s -
\frac{m_s\,m}{K} - a\,m_s\,m_i,\\ \frac{dm_i}{dt}
&= -c\,m_i - \frac{m_i\,m}{K} + a\,m_s\,m_i,
\end{align}
\end{subequations}
where the subscripts $i$ and $s$ refer to infected
and susceptible animals (in this case, mice) respectively,
$m=m_s+m_i$ is the total population, $b$ the birth
rate, $c$ the natural death rate, $K$ the
environmental parameter, and $a$ is the
transmission rate responsible for infection.

The fact, well-known to Hantavirus biologists
\cite{mills99b,kuenzi99}, that infection does not
affect the lifespan of the infected mice, is
naturally reflected in the mathematical
observation that the total population $m$ is
independent of all information regarding the
infection process.   The total population $m$
obeys a logistic equation, whose solution is
known.  Noticing this, the solution of system
(\ref{eq:ak}) can be easily obtained analytically
as shown by one of the present authors in Ref.
\cite{vmkpasi}.

This suggests that we first recast the
Abramson-Kenkre (AK) equations (\ref{eq:ak}),
changing the variables $m_i$ and $m_s$ to the
total population $m$ and the infected fraction
$\chi \equiv m_i /m$.   It is straightforward to
write
\begin{subequations} \label{eq:new_ak}
\begin{align} \frac{dm}{dt} &= (b-c)\,m
-\frac{m^2}{K},\\ \frac{d\chi}{dt}
&=-b\,\chi+a\,m\,\chi\,(1-\chi).
\label{eq:direct_infection} \end{align}
\end{subequations}
The first of these equations merely describes the
logistic evolution of the total population. The
second equation has an interesting structure. The
two terms on the right side have opposite
signs.  The first term, $-b\chi$, plays a role
against infection because birth of new individuals
always decreases the infected fraction $\chi$:
 the offspring are always susceptible (not
infected). The second term, $a m \chi (1-\chi)$,
represents the flux of individuals from
susceptible to infected. This flux occurs as a
result of transmission of infection between a
susceptible individual and an infected one by
direct contact. The transmission is represented by
the product of the infected fraction, the
susceptible fraction, and, of course, the total
population.

\begin{figure}
  \includegraphics[height=4in, width=6in]{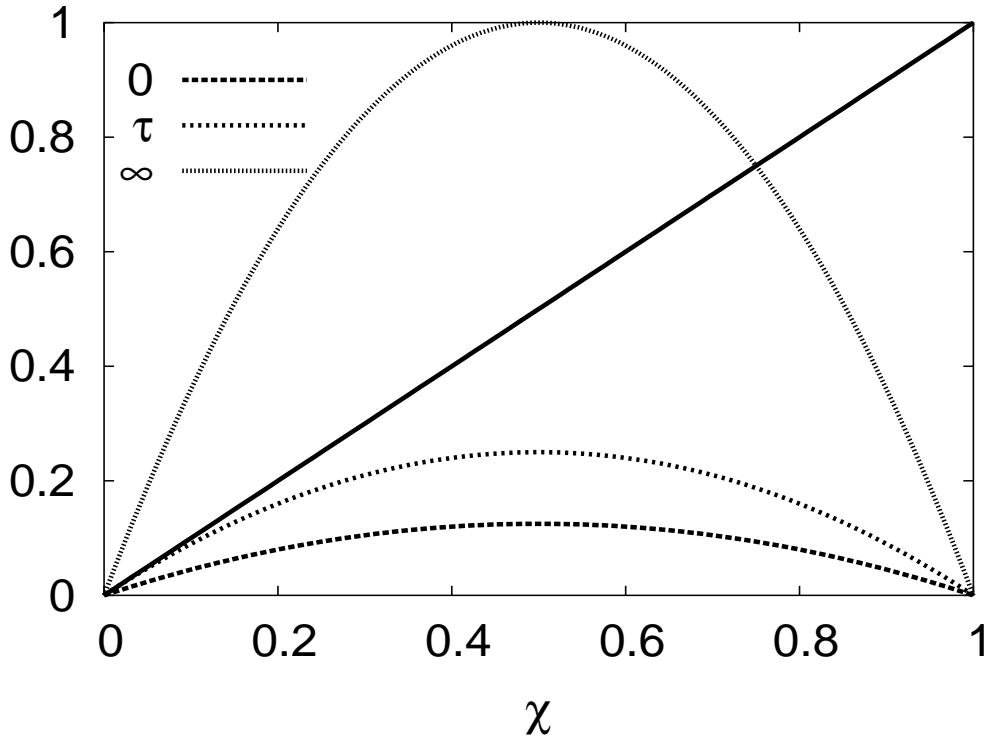}
  \caption{Balance between the two terms in the infection evolution.
  The straight line has slope $b$ and represents the decrease of the infected fraction due to birth of susceptible individuals. The parabolas represent the transfer of infection and their strength is proportional to the total population $m$. Three cases are shown. The lowermost parabola describes the initial time $t=0$ at which no nonzero intersection exists. The topmost parabola describes the eventual situation at  $t=\infty$, i.e., the steady state. The central parabola for which the straight line is a tangent describes the time at which the nontrivial intersection appears. This is the time $\tau$ at which the infection turns from its tendency to vanish and begins to rise to the nonzero steady state value.}\label{fig:balance}
\end{figure}

\begin{figure}
  \includegraphics[height=4in, width=6in]{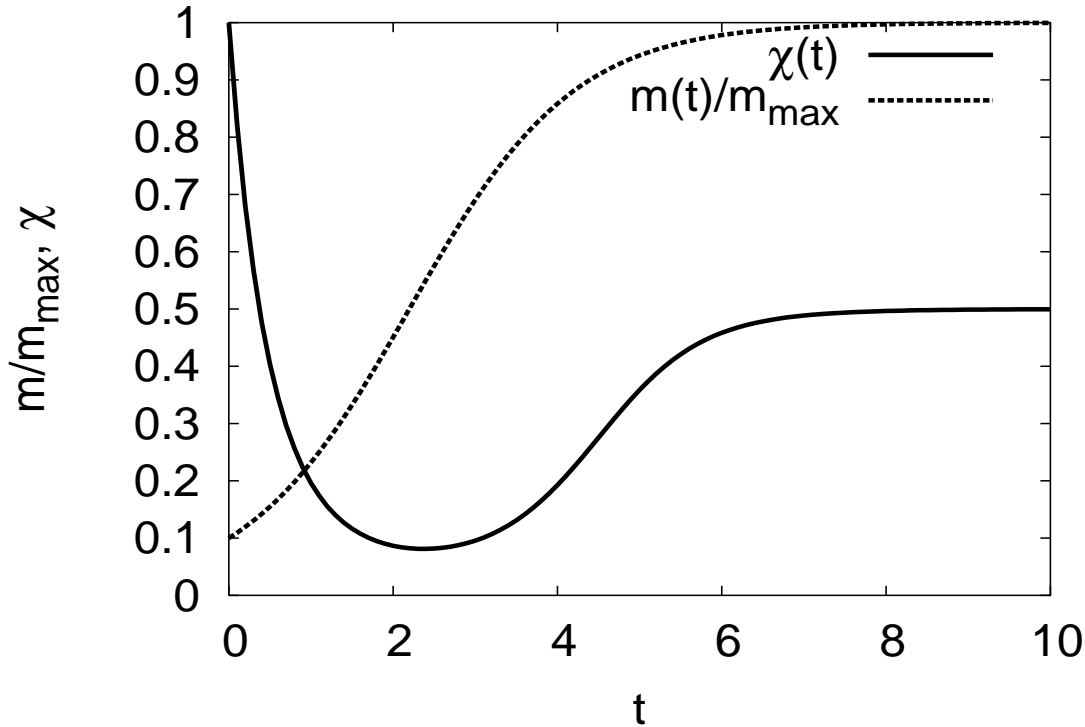}
  \caption{Time dependence of the infected fraction  $\chi$ and the total population $m$ corresponding to the situations depicted in Fig. \ref{fig:balance}. Parameters and initial conditions are arbitrary:
  $a=4,b=2,c=1,K=1$, $m(0)=0.1$ and
  $\chi(0)=1$. We see that  $\chi$ first tends to  vanish and then turns to its eventual steady state value. Time $t$ is plotted in  units of $\frac{1}{b-c}$. }
  \label{fig:caida}
\end{figure}

The system (\ref{eq:new_ak}) has four equilibria.
Two of them are irrelevant because one is the null
state and the other has negative population for
all values of the parameters.   With the help of a
linear stability analysis
\cite{bib:stability,bib:stability2} of the
equilibria, it can be shown that the other two
equilibria interchange their stability character
at a critical value of the parameter set. The
state with infection different from zero
($\chi>0$) is stable only if, as given in
Ref.\cite{ak},
\begin{gather} \label{eq:condicion}
K \, (b-c) \,> \,b/a.
\end{gather}
If this condition is not fulfilled, then the
steady state has no infection, i.e. $\chi=0$.
Equation (\ref{eq:direct_infection}) also suggests
an intuitive graphical procedure to ascertain the
presence and magnitude of the infection. A plot of
the $\chi$-dependence of each of the terms,
$b\chi$ (straight line), and $am\chi(1-\chi)$
(inverted shifted parabola) indicates in the
steady state the presence (absence) of infection
if the two curves do (do not) intersect each other
at a $\chi$ value other than $0$. See Fig. \ref{fig:balance}.
As $m(t)$ evolves in time via its logistic equation, 
the term $am\chi(1-\chi)$ changes. If its initial and final values are small and large respectively, the nontrivial intersection of the curves (nonzero $\chi$) will be absent at first, but present later on. Fig. \ref{fig:balance} shows such a case. We have labeled the initial and final situations by $0$ and $\infty$ respectively, and the critical situation, wherein the nontrivial intersections just begins to appear, by the the time $\tau$ taken for it to occur. The time dependence of the infected fraction and the total population is plotted in Fig. \ref{fig:caida}. We notice the clear tendency of the infection first to disappear (corresponding to the fact that the nontrivial intersection does not yet exist in Fig. \ref{fig:balance}) followed by evolution to the eventual steady-state value (corresponding to the intersection with the top curve in Fig. \ref{fig:balance}).

\begin{figure}
  \includegraphics[height=4in, width=6in]{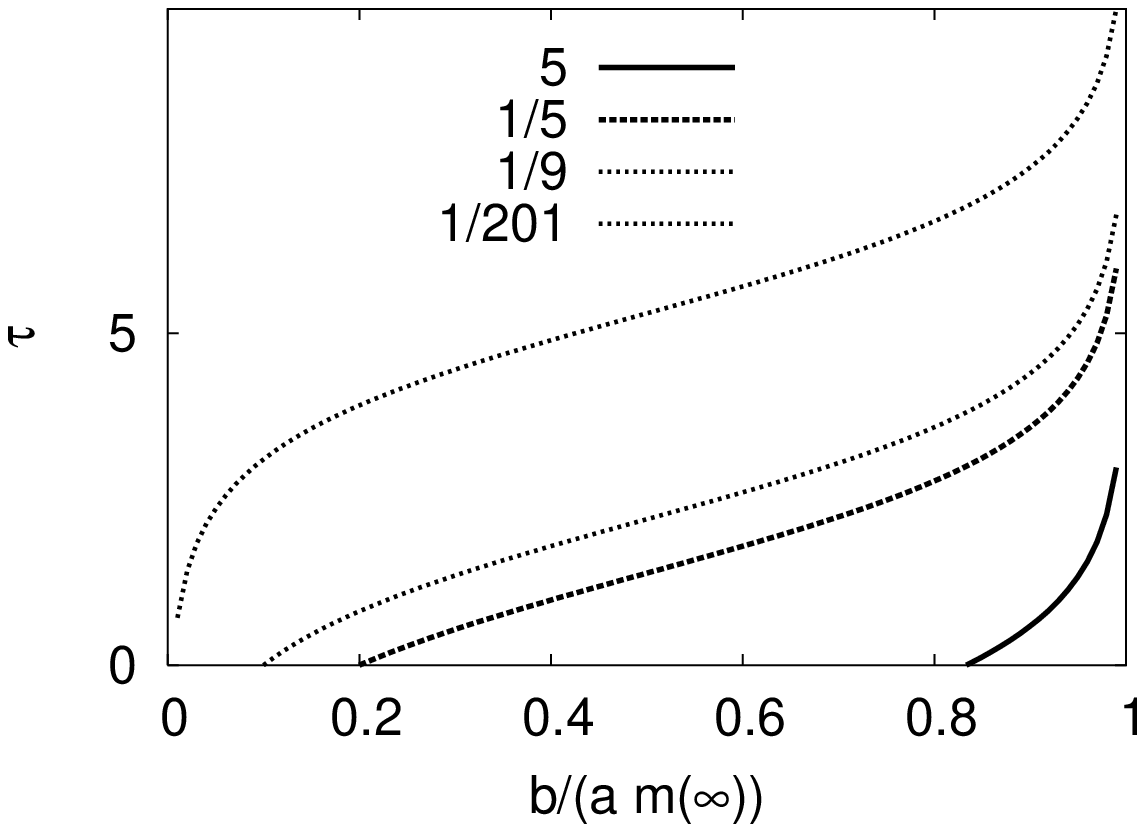}
  \caption{The delay time $\tau$ taken by the infected fraction to change its vanishing tendency and turn towards the nonzero steady state value plotted as a function of the ratio of the opposing rates $b$ and $a m(\infty)$. In this simple case $\tau$ is given by the analytic expression $(b-c)\tau = \ln\left( \frac{1-m(\infty)/m(0)}{1-m(\infty)/m_{c}}
\right)$ where $m_{c}= b/a$ is the critical carrying capacity. To be noted are the initial rise of $\tau$, the point of inflection and the blow up at the right extreme which  signifies that a nonzero infected fraction cannot be supported at higher values of the abscissa. The delay time is plotted in units of $\frac{1}{b-c}$.}
  \label{fig:delaydirecto}
\end{figure}

The delay $\tau$ in the onset of infection is plotted in Fig. \ref{fig:delaydirecto} versus the dimensionless ratio of the two rates that enter into the balance as clear from equation (2b). An initial increase, a point of inflection, and an eventual blow-up at the point the rate ratio equals $1$, are to be noted in the $\tau$ curve. The blow-up signals that the nontrivial intersection is always absent: the steady-state infection vanishes.

\subsection{Incorporating cross-infection}

Unlike with the Hantavirus, the spread of the West Nile virus
requires the presence of both mosquitoes
and birds:
the West Nile virus is transmitted through cross-infection.
This means that an infected individual infects a
susceptible individual of the \emph{other}
taxon. A mosquito infects a bird and vice-versa.

Therefore, in place of the AK equations
(\ref{eq:ak}), we write
\begin{subequations} \label{eq:cross_infection_old}
\begin{align}
\frac{d m_s}{dt} &= b\,m - c\,m_s -
        \frac{m_s\,m}{K} - a\,m_s\,A_i ,
        \label{eq:ci_ms} \\
\frac{dm_i}{dt} &=-c\,m_i - \frac{m_i\,m}{K} +
a\,m_s\,A_i , \label{eq:ci_mi} \\
\frac{d A_s}{dt} &= \beta\,A -\gamma\,A_s
        -\frac{A_s\,A}{\kappa} -
        \alpha\,A_s\,m_i ,
        \label{eq:ci_as}\\
\frac{d A_i}{dt} &= -\gamma\,A_i
        - \frac{A_i\,A}{\kappa}
        +\alpha\,A_s\,m_i,
        \label{eq:ci_ai}
\end{align}
\end{subequations}
where the subscripts $i$ and $s$ refer, as before,
to the infected and susceptible state
respectively.   The symbol $m$ now represents
mosquitoes rather than mice. The symbol $A$ (after Latin \emph{avis})
represents birds with $\beta$ as the birth rate,
$\gamma$ as the death rate, and $\kappa$ as the
environmental parameter. The \mbox{cross-infection}
rates are $a$ and $\alpha$. Equations
(\ref{eq:cross_infection_old}) are formally
symmetric in the two taxa. In each case,
infected as well as susceptible individuals breed
susceptible individuals of their own species. Also, in each case,
infected individuals infect susceptible members of
the \emph{other} species.  In addition, each
species has its own vital dynamics---each modeled by a
logistic equation---via its own birth rate, death
rate, and environmental parameter. For the sake of
explanation let us call one of the taxa
(i.e., the birds) the host population and the
other (i.e., the mosquitoes) the vector. An
infected individual of the host population, $A_i$,
transmits the disease to a susceptible member of
the vector taxon, $m_s$.  This member becomes
infected, increasing the infected population of the vector taxon,
$m_i$.
Only then is this newly
infected individual
able to transmit the disease to a susceptible member of
the original host population, $A_s$.   As a result of
the last interaction, an individual of the host
species will finally join the infected population $A_i$. There is
thus an underlying cyclic process. In this way,
the infection process can be thought of in two
stages.   One is the acquisition of the infection
by the `vector'.  The other is the transmission
to the `host' population. Each of these requires
direct contact between transmitter and receptor.
Therefore, the magnitude of each infection process
depends on the number of receptors, the number of
transmitters, and the respective infection rates.

How is the equation set (\ref{eq:new_ak}) augmented by the
incorporation of cross-infection? To answer this we rewrite
equations (\ref{eq:cross_infection_old}) in terms of total
populations and infected fractions
\begin{subequations} \label{eq:cross_infection}
\begin{align}
\frac{dm}{d t} &= (b-c)\,m -\frac{m^2}{K}, \\
\frac{d\chi_m}{dt} &= -b\,\chi_m
+ a\,A\,\chi_A\,(1-\chi_m) , \label{eq:main_prop}\\
\frac{dA}{dt} &= (\beta-\gamma) A - \frac{A^2}{\kappa} ,\\
\frac{d\chi_A}{dt} &=-\beta\, \chi_A
        + \alpha\,m\,\chi_m\,(1-\chi_A).
\label{eq:vector_prop}
\end{align}
\end{subequations}
We see that the evolution of the total population
of the mosquitoes, $m=m_i +m_s$, and of the birds,
$A=A_i +A_s$, is formally unchanged. The
respective infected fractions $\chi_m=m_i / m$ and
$\chi_A=A_i / A$ clearly show the effect of
cross-infection. As in equations
(\ref{eq:new_ak}), we see that the infected
fraction of either taxon decreases as the result
of births in that taxon because only susceptible
individuals are born (represented by the terms $-b
\chi_m$ and $ -\beta \chi_A$ in
(\ref{eq:main_prop}) and (\ref{eq:vector_prop})
respectively).

Linear stability analysis
\cite{bib:stability,bib:stability2} of
(\ref{eq:cross_infection}) along
the lines of (\ref{eq:ak}) shows that the equilibria when
infection is different from zero
($\chi_m\neq0,\chi_A \neq 0$) are stable
only when
\begin{align}
\label{eq:condicion2}
K(b-c) \kappa (\beta-\gamma)
> (b/a) (\beta/\alpha).
\end{align}
Equation (\ref{eq:condicion2}) represents a
generalization of (\ref{eq:condicion}) to
the cross-infection case.  Threshold values for
infection survival depend on products of
quantities characteristic of the two taxa.

\section{Incorporating Realistic Features of the
West Nile Virus}\label{realistic_model}

The preceding analysis has focused on the
consequence of replacing same-taxon infection by
cross-infection typical of West Nile virus and has
relied on a highly simplified and symmetrical
model. We now include three realistic features of
the West Nile virus: (i) the possibility of vertical
transmission in mosquitoes, (ii) the possibility of
infection-caused mortality of birds, and (iii) time scale disparity between
mosquitoes and birds.

\subsection{Partial Heritage of the Infection:
Vertical Transmission}

Vertical transmission has been strongly suspected
in the West Nile virus \cite{Nasci}. By this term
is meant the passage of virus from
infected individuals to their offspring via the
process of birth. Therefore, we now consider that
some of the offspring of infected mosquitoes are
infected during the formation of eggs.  We take the rate
of mosquitoes being ``born" (hatched) already infected as
$b_i$. Infected mosquitoes can only be born from
infected mosquitoes. The total rate of
mosquitoes born from infected ones is still $b$.
To incorporate this effect, we subtract the term
$b_i m_i$ from equation~(\ref{eq:ci_ms}) and add it to the
right-hand side of (\ref{eq:ci_mi}).

While vertical transmission could lead to the survival of infection within mosquito larvae and reemerge in the spring, within the framework of equations we have adopted here, this modification has no important qualitative effect if $b_i
\neq b$, i.e, if not all offspring of infected
mosquitoes are hatched infected.
It only changes the critical values and
the evolution times in a straightforward way.
Linear stability analysis
\cite{bib:stability,bib:stability2} shows that now the
condition for the steady infected state is
\begin{gather} \label{eq:condicion3}
K (b-c) \, \kappa (\beta-\gamma) \, > \, \left(\frac{b-b_i}{a}\right)
\left(\frac{\beta}{\alpha}\right).
\end{gather}
The asymmetry between the susceptible and infected populations,
which favors the former as in Eq.~(\ref{eq:condicion2}), still
holds provided $b_i\neq b$. The modification in the bifurcation
point is changed only quantitatively.

\subsection{Mortality Increase due to Infection}

By contrast to vertical transmission discussed
above, the increase of the mortality rate in the
bird population due to infection can have
substantial consequences within our analytical framework.  In addition to the
vertical transmission modifications involving
$b_i$, we now replace $\gamma A_i$ in
equation~(\ref{eq:ci_ai}) by $(\gamma+\delta) A_i$
where $\delta$ represents the infection-based
contribution to the bird mortality rate. The
generalization of equations
(\ref{eq:cross_infection}) is now
\begin{subequations} \label{eq:final_system}
\begin{align} \frac{dm}{dt} &=    (b-c)\,m   -
\frac{m^2}{K} , \label{eq:total_modif}\\
\frac{d\chi_m}{dt} &= - (b-b_i)\,\chi_m
+a\,A\,\chi_A\,(1-\chi_m) , \label{eq:evochi}\\
\frac{dA}{dt} &= (\beta-\gamma-\delta \chi_A)\,A -
\frac{A^2}{\kappa} \\ \frac{d\chi_A}{d t} &= -
\beta\,\chi_A +
(\alpha\,m\,\chi_m\,-\delta\,\chi_A)(1-\chi_A).
\label{eq:new_asymetri} \end{align}
\end{subequations}

The condition for the stability of the state
with nonzero infection is,
\begin{equation} \label{eq:condicion4}
K\,(b-c)\,\kappa\,(\beta-\gamma) \, >
\,\left(\frac{b-b_i}{a}\right)\left(\frac{\beta+\delta}{\alpha}\right).  \end{equation}
The difference with the previous condition
(\ref{eq:condicion3}) for linear stability of the
infected state is the replacement of $\beta$ in
the right side by $\beta+\delta$.
This result may appear puzzling. That a death rate
contribution $\delta$ should \textit{add} to,
rather than subtract from, the birth rate $\beta$
may look counterintuitive. However, it arises simply from
the fact that the relative increase of the number of susceptible
over infected birds is a consequence of increased infected
mortality as well as of the birth process for susceptible birds.

\begin{figure}[!htbp]
\includegraphics[height=4in, width=6in]{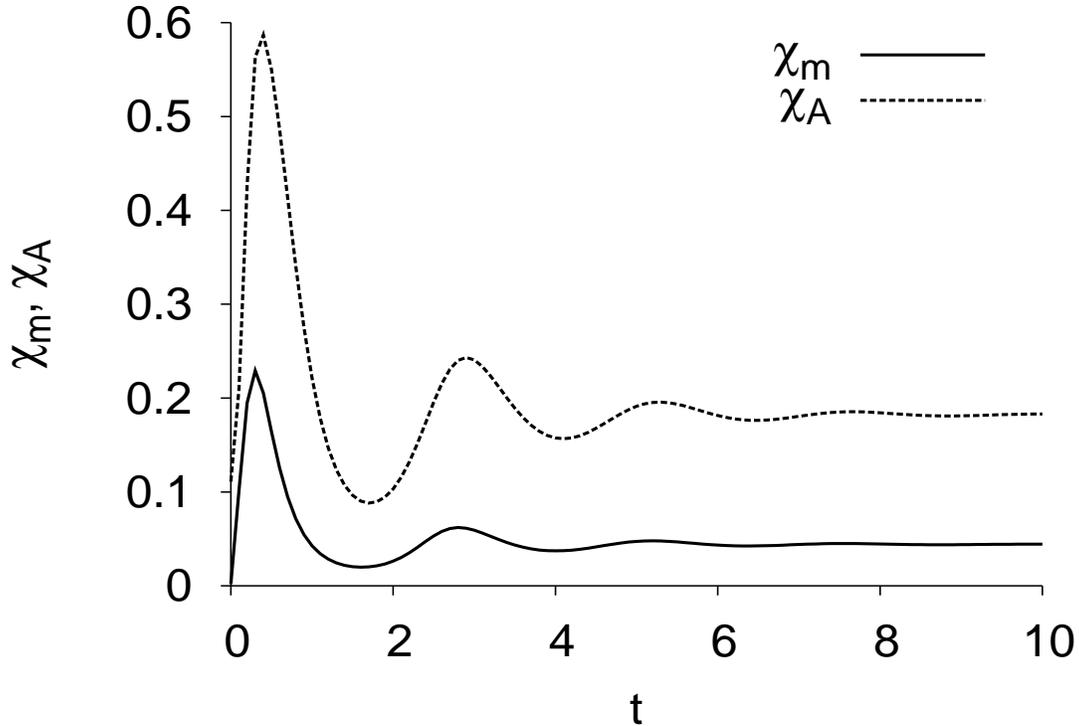}
\caption{Damped oscillations in the infected
fractions during their approach to steady state, arising from the
presence of the extra bird mortality (the additional death rate is $\delta$) from infection. Time is plotted in units of $1/(b-c)$. Parameters are arbitrary:
$b=20,c=19,b_i=0,a=0.25,\beta=5,\gamma=3.5,\alpha=1,\delta=6,
K=50,\kappa=50$} \label{fig:damping}
\end{figure}

A noteworthy outcome of the increase of mortality
in birds due to infection is damped oscillations
in  $A,\chi_m,\chi_A$ as they approach their
steady values.
The mosquito population $m$ is not affected.
We show in Fig. \ref{fig:damping} this damped
approach  to equilibrium of the
infected fractions $\chi_m$ and $\chi_A$.

The damped oscillations behavior does not require cross-infection
as long as the extra death rate of infected individuals is
present. Indeed, the phenomenon can beunderstood more easily in the
simpler case of direct infection. Consider for this case
\begin{subequations}
\begin{align}
\frac{dA}{dt} &= (\beta-\gamma-\delta
\chi)\,A-\frac{A^2}{\kappa} \label{eq:damp_eqA}\\
\frac{d\chi}{dt} &= - \beta\,\chi +
(\alpha\,A\,-\delta)\,\chi\,(1-\chi).
\label{eq:damp_eq}
\end{align}
\end{subequations}
Whereas the coefficient of the negative term $-\beta \chi$ in
equation (\ref{eq:damp_eq}) is fixed for all time $t$, the
transmission term is time dependent because of the time dependence
of $A$. That dependence is influenced, as (\ref{eq:damp_eqA})
shows, by the evolution of the infected fraction $\chi$ through
the extra death rate via the term ($- \delta \chi A$). Suppose
that we are in a state in which the infected population is
growing. While the amount of infected individuals increases, the
number of deaths in population $A$ increases. Thus, $A$ decreases.
If the decay of the value of the total population $A$ is such that
$(\alpha\,A-\delta) < \beta$ at some moment, the system will be in
a situation wherein the infection is more likely to disappear than
not. So, the infection decreases and the population recovers. This
might be repeated several times until the system reaches the
steady state. If, however, the extra death rate is high enough,
oscillations might be completely suppressed and the infection may go directly to a
steady state or disappear entirely.

\begin{figure}[!htbp]
\includegraphics[height=4in, width=6in]{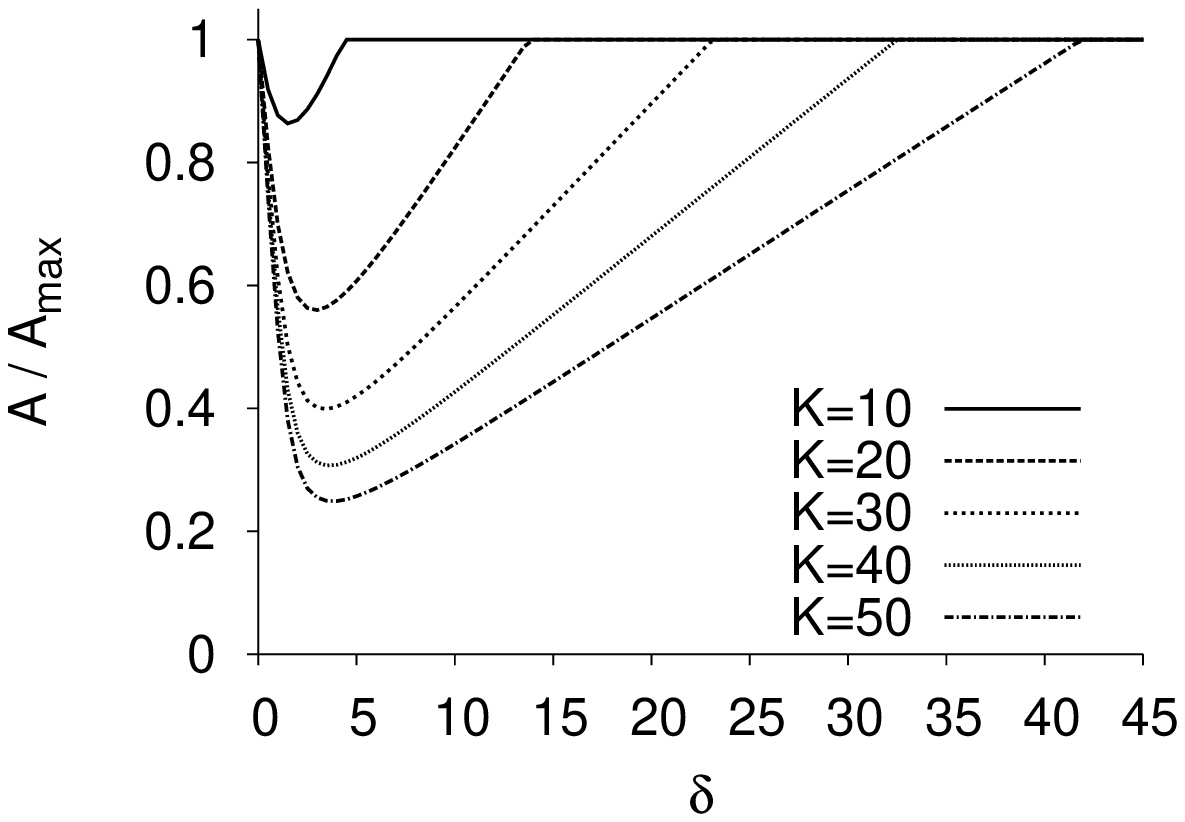}
\includegraphics[height=4in, width=6in]{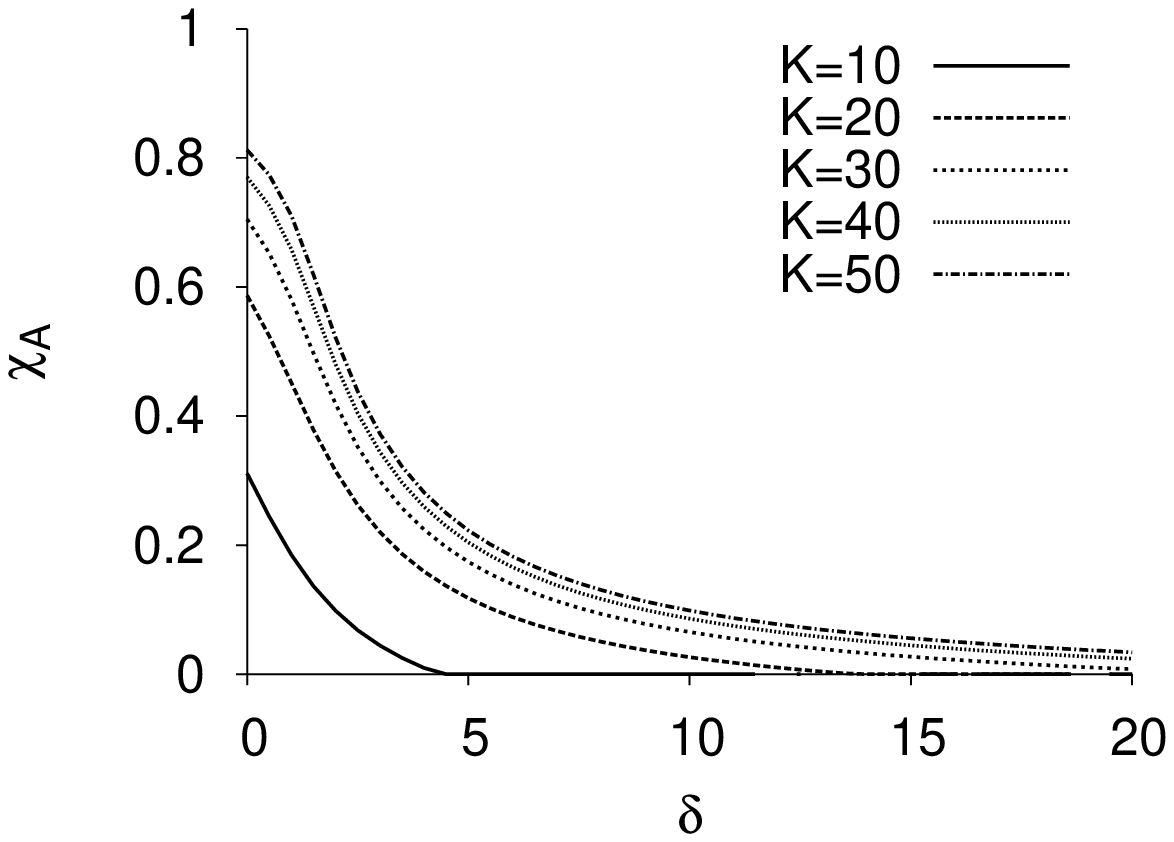}
\caption{Effect of the increased mortality
rate $\delta$ on the steady state values of the total
bird population and  the  infected fraction. The
total population is expressed relative to
the carrying capacity $A_{max}$ while the increased mortality
rate $\delta$ is expressed relative to the rate difference
$b-c$. Parameters are arbitrary:
$b=20,c=19,b_i=0,a=0.25,\beta=5,\gamma=3.5,\alpha=1,
\kappa=50$. The dependence of the infected fraction is monotonic while
that of the total population has a minimum.}
\label{fig:optimal}
\end{figure}

The oscillations arise, thus, from the need of the presence of
infection from transmission to occur, combined with the decrease
in the population due to mortality from infection. This
combination has another effect: the extra death rate parameter
$\delta$ has an optimal value (see Fig. \ref{fig:optimal}). If
$\delta$ is too large,  the infection is eliminated completely because
the indispensable elements for transmission (infected individuals)
are killed very quickly (see Fig. \ref{fig:optimal} B). Continuity 
of the infection requires a value for $\delta$ which provides the
proper balance between the infection process and the death
of infected birds.

\subsection{Time Scale Disparity}

\begin{figure}[!htbp]
\includegraphics[height=4in, width=6in]{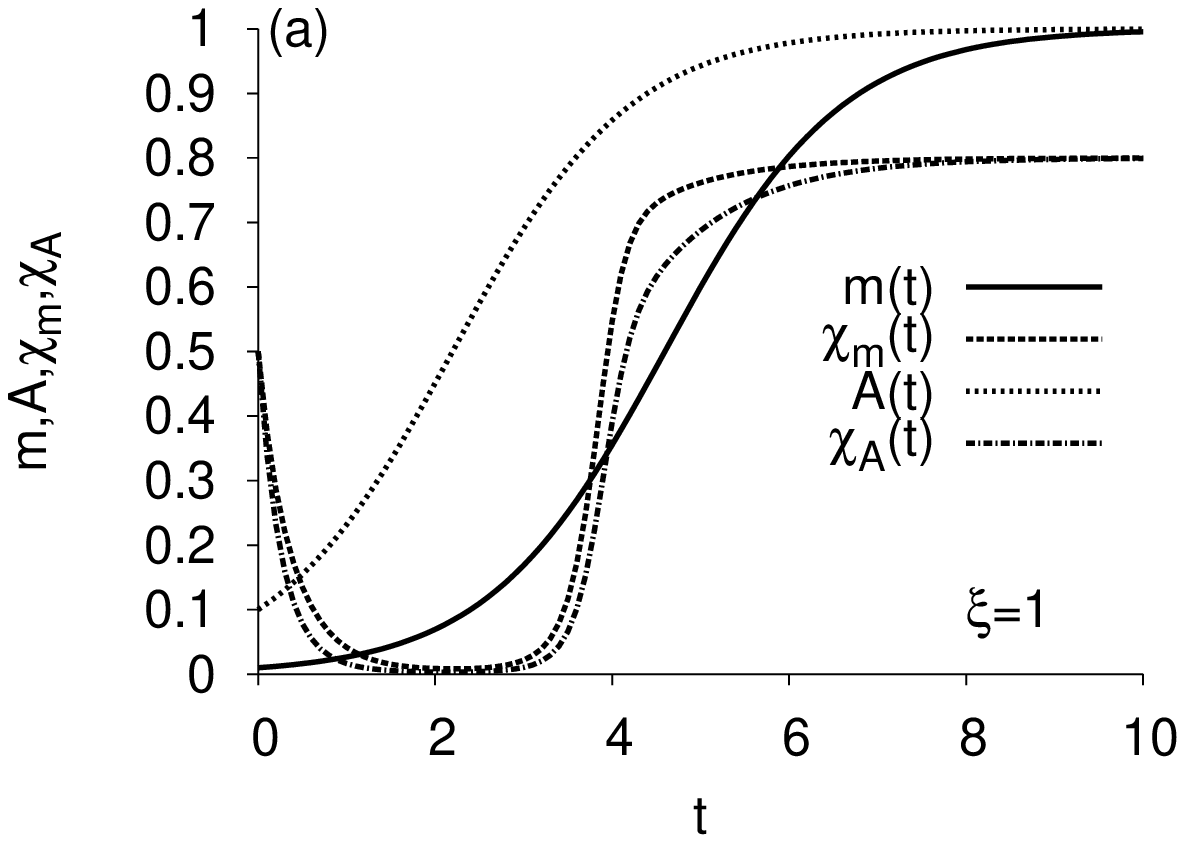}
\includegraphics[height=4in, width=6in]{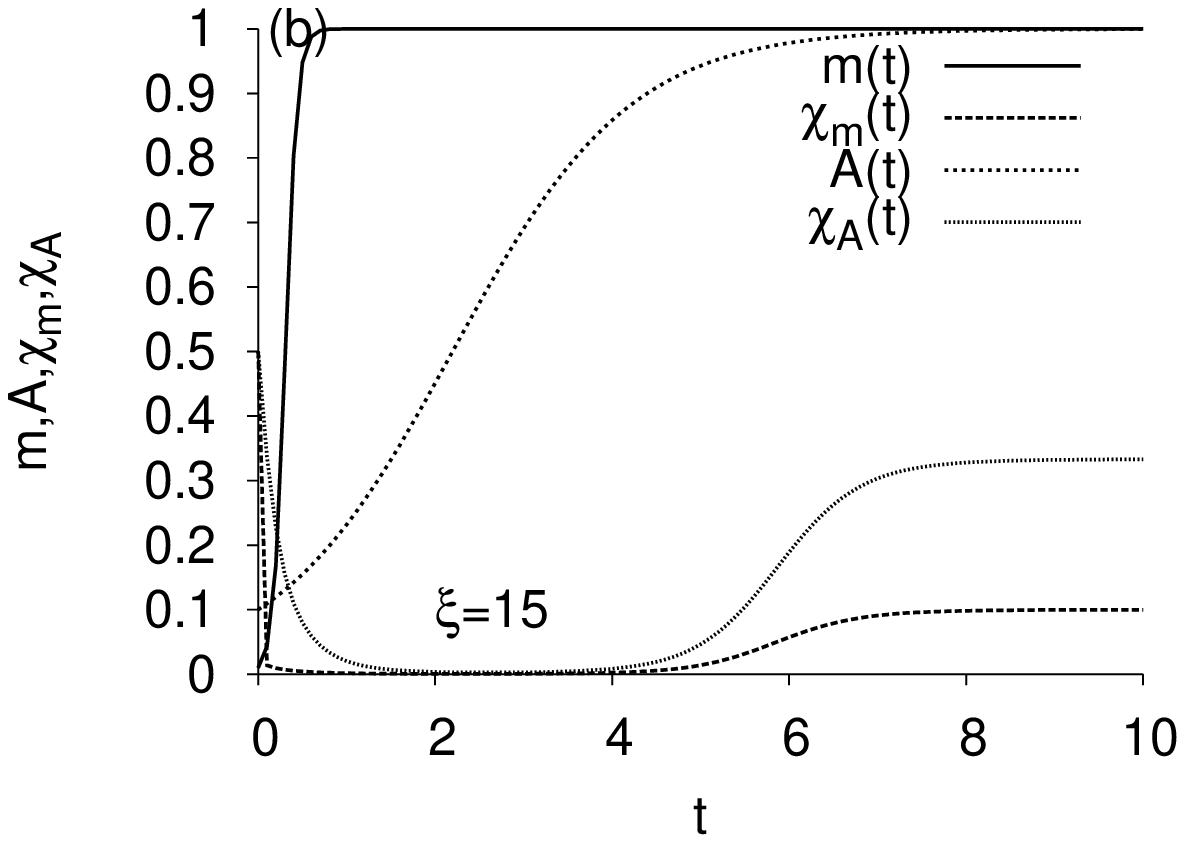}
\caption{Effect of time scale disparity on the time evolution of
the mosquito and bird populations $m$ and $A$, and their
respective infected fractions $\chi_m$ and $\chi_A$. Time is
plotted in units of the rate difference $1/(\beta - \gamma)$. The
disparity factor $\xi$ (see text) is $1$ in (a) and $15$ in (b).
This means that $b=\xi \beta$ and $c=\xi \gamma$. Other parameters
are arbitrary: $\beta=4,\gamma=3,\kappa=1,a=20,\alpha=20,
K=\frac{1}{b-c}$. The initial conditions have been taken to be
$A(0)=0.1, m(0)=0.01, \chi_m(0)= 0.5 ,\chi_A(0)=0.5$. }
\label{fig:thre}
\end{figure}

\begin{figure}[!htbp]
\includegraphics[height=4in, width=6in]{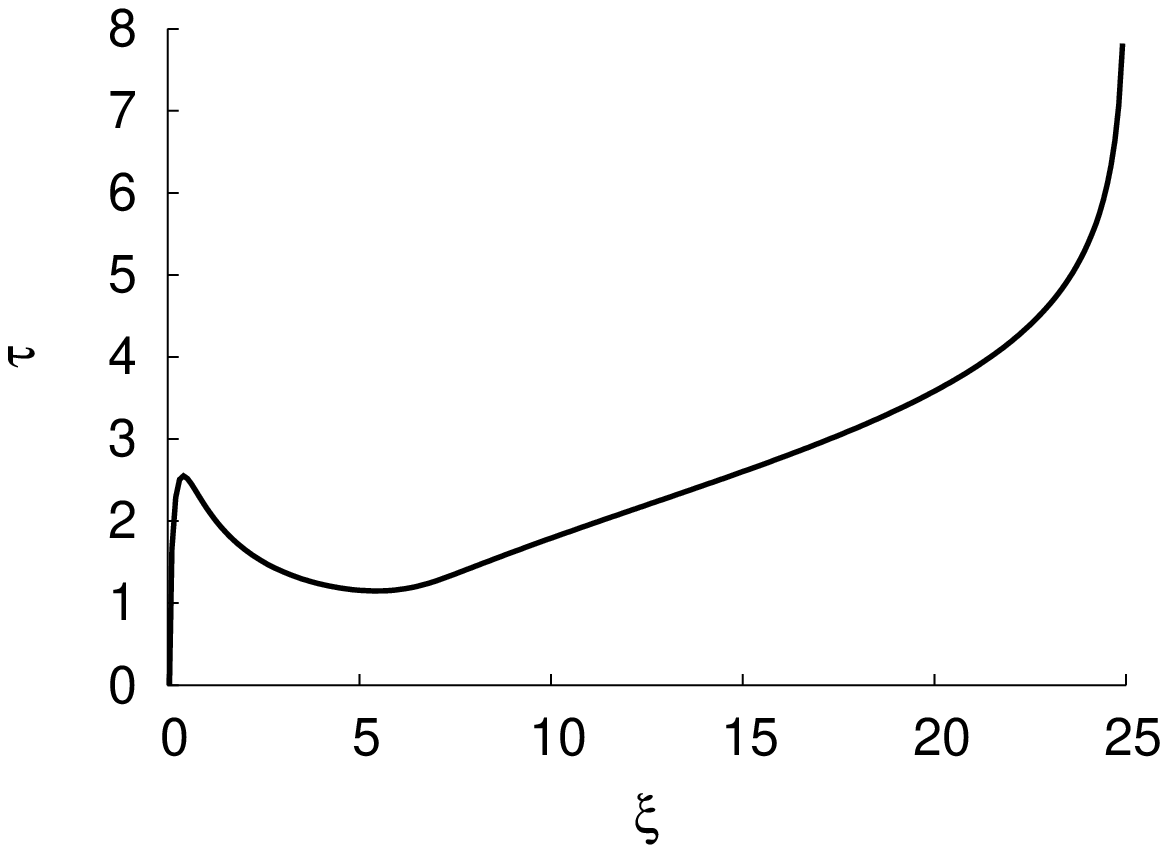}
\caption{Strongly nonmonotonic dependence of the delay $\tau$ (see text) in
the appearance of the infection as a function of disparity factor
$\xi$. The delay is plotted in units of the rate difference $1/(\beta
- \gamma)$. Parameters are as in Fig. \ref{fig:thre}. Sharply
different rates of increase of the delay with the disparity factor
are seen, along with a blow-up at the extreme right. The
blow-up occurs at the critical value of
$\xi_c=\frac{\beta}{\alpha m(\infty)} \frac{b}{a A(\infty)}$, beyond which the state with nonzero infection cannot be supported.} \label{fig:delay}
\end{figure}

A peculiarity of the West Nile virus epidemic is that
processes involving mosquitoes and birds occur on
quite different time scales. The natural lifespan
of a mosquito is not generally longer than a month while
that of a bird might be several years.  The
birth and death rates of one taxon are thus very
different from those of the other taxon.  In
order to understand what effect this might produce
on the dynamics of infection, we first make the
simple assumption that the corresponding rates for
the two taxa differ by a factor $\xi$, i.e.,
\begin{equation}
\xi = \frac{b}{\beta} = \frac{c}{\gamma}.
\end{equation}
We have followed several time scale separation schemes based on recasting the terms appearing in the evolution equations into those divided by $\xi$ and those independent of $\xi$. For large time scale disparity, the former drop out of the evolution effectively. Our studies of time scale disparity along these lines are of interest to the nonlinear dynamics of the system but appear so far to be of much less value to the understanding of the epidemic. Therefore, we display here only  the temporal evolution for two
different values of $\xi$ (see Fig. \ref{fig:thre}), and the corresponding interesting nonlinear dependence of the delay $\tau$ in the onset of steady state infection on $\xi$ (see Fig. \ref{fig:delay}).

Figure \ref{fig:delay} is the generalization of Fig. \ref{fig:delaydirecto} to the cross-infection situation present in the West Nile virus epidemic.
If the condition (\ref{eq:condicion2}) is
fulfilled, the variables $\chi_{m,A}$ attain their
limiting nonzero values as $t\rightarrow+\infty$.
However, as seen in Fig.~\ref{fig:thre}, for
some initial amount of time, they could appear to
be attracted to a state with no infection.  A
measure of this transient time can be estimated by
noticing that the condition (\ref{eq:condicion2})
for the stability of the state with infection can
be written as
\begin{equation}
\lim_{t \to \infty} A(t) m(t) > (b/a) (\beta/\alpha),
\end{equation}
and asking for the value $\tau$ of the time $t$ at
which this condition starts to be satisfied by the
system. In other words, a measure of the delay
$\tau$ can be obtained from
\begin{equation}
A(\tau) m(\tau) = (b/a) (\beta/\alpha).
\end{equation}
Because the functions $A(t)$ and $m(t)$ are known
solutions of the logistic equation, we can solve
this equation numerically. In this way, we can
obtain the delay time $\tau$ as a function of
$\xi$ for any initial conditions and parameter
values.  An example of the dependence of the delay
time $\tau$ on the time scale disparity factor
$\xi$ is shown in Fig.~\ref{fig:delay}. In
changing $\xi$ we keep the bird parameters
constant and change only the mosquitoe rates. Additionally, the environmental parameter of the
mosquitoes, $K$, is reduced by the same factor as
$\xi$ is increased, in order to keep the carrying
capacity $K(b-c)$ constant.

The value of $\tau$ plotted in Fig. \ref{fig:delaydirecto} is also obtained in this fashion although only a single population (ratio) enters in that case into the left (right) hand side of the criterion equation. The blow-up feature at the limiting values of the rate ratio is common to both the direct and the cross infection cases (Fig. \ref{fig:delaydirecto} and Fig. \ref{fig:delay}). The strong non-monotonic behaviour  of $\tau$ in the cross-infection case is interesting and merits further study. 

Further time scale disparity conclusions cannot be drawn until the relative values of the cross-infection parameters are known. Thus, it is possible to recast the equations of the mosquito and bird populations in dimensionless form to make clear the time scale disparity obvious from the numerical solutions displayed in the figures. Defining $\mu=m/m_{max}$ and ${\cal A}=A/A_{max}$, where $m_{max}$ is the mosquito carrying capacity $K(b-c)$ and  $A_{max}$ is the bird carrying capacity $\kappa(\beta-\gamma)$, evolution equations for the total populations in terms of the dimensionless time $t'=t(b-c)$ take the form
\begin{subequations}
\begin{align}
\frac{d\mu}{dt'} &= \mu-\mu^2 \label{eq:mu}\\
 \frac{d{\cal A}}{dt'} &= \frac{1}{\xi}({\cal A}-{\cal A}^2)
\label{eq:calA}
\end{align}
\end{subequations}
The time rate of change of the bird population is clearly slower than of the mosquito population by the disparity factor. This agrees with the quick rise of the mosquito population displayed in the figures. However, whether the infection ratios change on the same or disparate time scales depends on the relative values of the parameters $\frac{aA_{max}}{b-c}$ and $\frac{\alpha m_{max}}{\beta-\gamma}$. If they are of the same order as each other, bird infection will involve on a slower time scale than mosquito infection. For values we have taken to draw the plots displayed, both infected fractions appear to evolve on the same time scale.

There is another time scale disparity comment worth making. The short-time part of any logistic evolution is exponential increase. Observation times of interest in the kind of West Nile virus studies we have discussed in this paper are typically short on bird time scales. They are, however, not short on mosquito time scales. Therefore, the total bird population $A(t)$ might be taken to be $A(0)e^{(\beta-\gamma)t}$ to a good aprroximnation. The total mosquito population $m(t)$ should not be approximated in this manner.

\section{Conclusions} \label{conclusion}

In this paper we have presented some essential
features of a theoretical framework to analyze the
spread of the West Nile virus epidemic.  Explaining existing data has not been our aim because such data are rather scarce. Starting with
the AK equations given earlier for Hantavirus
investigations \cite{ak}, we have developed the
West Nile virus theory in three stages.

First, we modified the Hantavirus equations
(\ref{eq:ak}) by replacing same-taxon infection
by cross-infection, peculiar to the West Nile virus.  At
this stage, we retained Hantavirus features by
assuming that all
susceptible organisms have comparable lifespans,
no organisms are born infected, and the infection
does not affect the death rate of members of
either taxon.
Our results (equations
\ref{eq:cross_infection_old} and
\ref{eq:cross_infection}) showed that the condition
for steady-state infection is a generalization
from the one obtained for the
Hantavirus (no cross-infection), involving a combination of the
parameter of the two taxa.

In the second stage, we studied realistic features of the West
Nile virus by including an analysis of vertical transmission for the mosquitoes, and
increased mortality rate in birds due to infection. We found that
vertical transmission does not affect the qualitative behavior of
the system within the framework of equations we have adopted. However, we found that damped oscillations in the evolution
emerge from the increased mortality rate in birds due to
infection. Furthermore, we saw that, while the dependence of the
infected fraction of birds on the increased mortality rate is
monotonic, the dependence of the total bird population is not,
there being a characteristic value at which the maximum number of
birds are killed. For lower as well as higher values of the
mortality rate, the bird population is larger. Finally, we found that disparate lifespans of mosquitoes and birds lead to the effect that the
delay in the onset of steady-state of infection depends
nonlinearly on the ratio of the characteristic times of the two
taxa.

The third stage of our investigations addresses an important feature of the West Nile virus epidemic: the
\textit{movement} of mosquitoes and
birds, particularly the \textit{migration} of
birds. Because this stage has not been completed, we  have not presented our results in this paper. However we state here the basic idea and the equations we use for this purpose. The equations are
\begin{subequations} \label{eq:super}
\begin{align} \frac{\partial m_s}{\partial t} &=
(b-c) m_s + (b-b_i) m_i - a m_s A_i - \frac{m_s
m}{K} + D_m \frac{\partial^2 m_s}{\partial x^2},
\nonumber \\ \frac{\partial m_i}{\partial t} &=
(b_i-c) m_i + a m_s A_i - \frac{m_i m}{K} + D_m
\frac{\partial^2 m_i}{\partial x^2}, \nonumber \\
\frac{\partial A_s}{\partial t} &= \beta A -\gamma
A_s - \alpha A_s m_i - \frac{A_s A}{\kappa} +\int
dy f(x,y) A_s(y,t) \nonumber ,\\ \frac{\partial
A_i}{\partial t} &= -(\gamma+\delta) A_i + \alpha
A_s m_i - \frac{A_i A}{\kappa} +\int dy f(x,y)
A_i(y,t). \nonumber \end{align}
\end{subequations}
The movement of the mosquitoes is considered diffusive and
represented by the diffusion constant $D_m$ while the long range
movement of birds (including, particularly, migration) is represented
by the integral terms involving $f(x,y)$. Information about the speed at which mosquitoes move, as well as their effectively enhanced mobility due to wind and related effects, is fed into $D_m$. An alternative to the integral description of the long-range motion of birds given above is a treatment through a partially systematic and partially stochastic term representing the appearance and disappearance of birds (and infection) at the site under investigation, as a result of their migration. Another important feature missing from the work reported in the present paper is the seasonal disappearance of mosquitoes, an essential part of the conduit of infection, when the temperature drops below that capable of sustaining them. Yet another is the possible of reemergence of infection in the spring from infected larvae. Work on all these aspects is under way
and will be reported in a future publication.
An alternative approach to the theory of the West
Nile virus based on a difference equation model \cite{thomas} that has appeared recently in the literature, has been brought to our attention.
In future work we will report similarities,
differences and domains of applicability of the
two formalisms.

\begin{acknowledgments}
One of the authors (VMK) thanks Marcelo
Kuperman for introducing him to the West Nile virus problem. 
IDP acknowledges the hospitality of the
University of New Mexico.  This work was supported in part by the
National Science Foundation via grant nos.  DMR-0097204, INT-0336343, by
DARPA-N00014-03-1-0900, and by a contract from Los Alamos National Laboratory
to the University of New Mexico (Consortium of the Americas for
Interdisciplinary Science).
\end{acknowledgments}

\end{document}